\documentclass{article} %
\usepackage{colm2024_conference}

\usepackage{booktabs}
\usepackage{graphicx}
\usepackage{enumitem}
\usepackage{wrapfig}
\usepackage{algorithm}
\usepackage{algpseudocode}

\usepackage{CJKutf8}
\usepackage{microtype}
\usepackage{amsmath}
\usepackage{amsfonts}
\usepackage{colortbl}
\usepackage[utf8]{inputenc}
\usepackage[T1]{fontenc}
\definecolor{lightgray}{rgb}{0.9,0.9,0.9}
\usepackage{caption}
\usepackage{subcaption}
\usepackage{xcolor}
\usepackage{setspace}
\usepackage{url}
\usepackage{multirow}
\usepackage{colortbl}
\usepackage{tabularx}
\usepackage{xltabular}
\usepackage{blindtext}
\usepackage{pgfplots}
\pgfplotsset{compat=1.18} 
\usepackage{tikz}
\usetikzlibrary{er, positioning,bayesnet}
\usepackage{makecell}
\usepackage{tipa}
\usepackage{siunitx}
\usepackage{nicefrac}
\usepackage{tocloft}
\usepackage{listings}
\usepackage[raster,skins]{tcolorbox} %
\usepackage{xltabular}
\usepackage{adjustbox}
\usepackage{xurl}
\usepackage{multirow}
\usepackage{threeparttable}
\usepackage{xcolor}
\usepackage{amsmath,amsfonts,bm}

\def\eqref#1{equation~\ref{#1}}
\def\1{\bm{1}}

\DeclareMathAlphabet{\mathsfit}{\encodingdefault}{\sfdefault}{m}{sl}
\SetMathAlphabet{\mathsfit}{bold}{\encodingdefault}{\sfdefault}{bx}{n}

\begin{document}
\begin{figure}[htbp]
    \centering
    \includegraphics[width=0.2\textwidth]{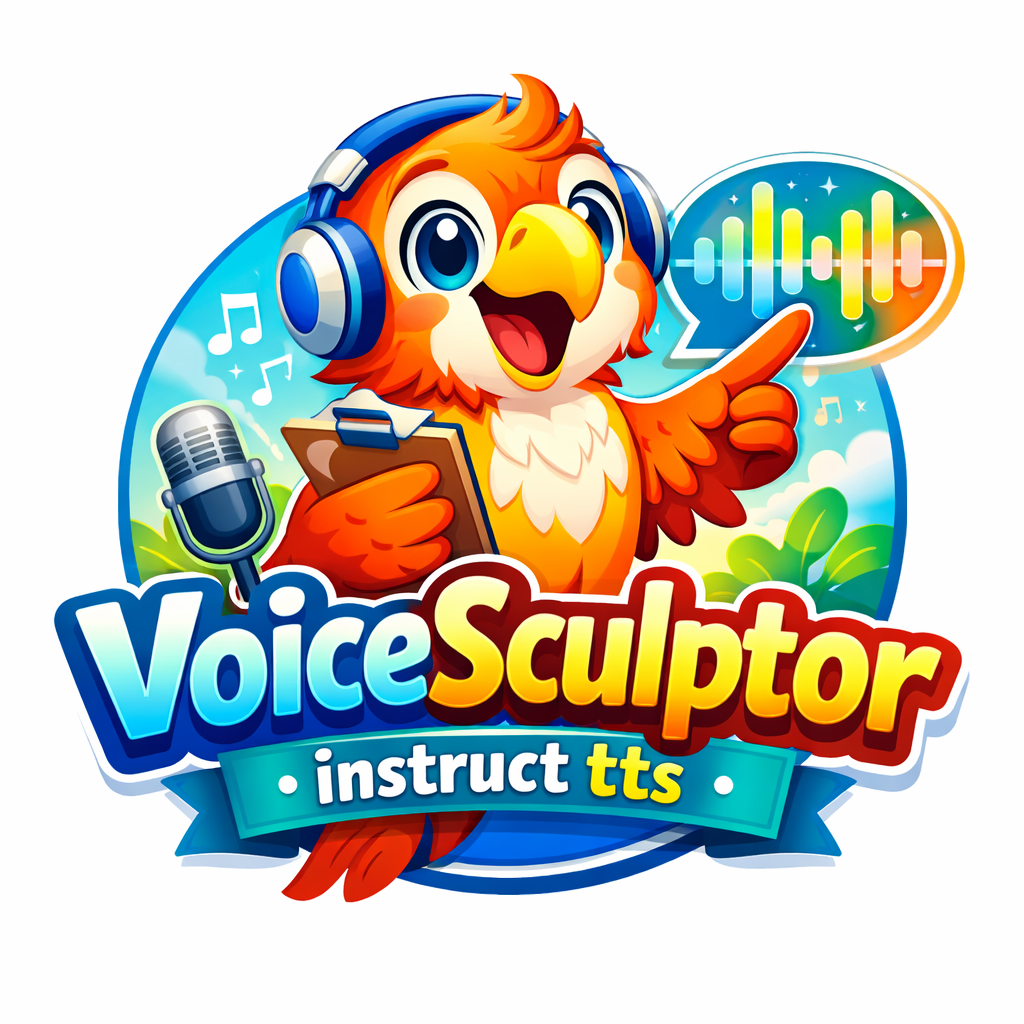}
    \label{fig:logo}
\end{figure}
\title{VoiceSculptor: Your Voice, Designed By You}

\author{
Jingbin Hu$^{1}$, Huakang Chen$^{1}$, Linhan Ma$^{1}$, Dake Guo$^{1}$, Qirui Zhan$^{1}$, Wenhao Li$^{1}$, Haoyu Zhang$^{1}$,
Kangxiang Xia$^{1}$, Ziyu Zhang$^{1}$, Wenjie Tian$^{1}$, Chengyou Wang$^{1}$, Jinrui Liang$^{1}$, Shuhan Guo$^{1}$, Zihang Yang$^{1}$,
Bengu Wu$^{2}$, Binbin Zhang$^{4}$, Pengcheng Zhu$^{1,4}$, Pengyuan Xie$^{3}$, Chuan Xie$^{3}$, Qiang Zhang$^{3}$, Jie Liu$^{3}$, Lei Xie$^{1}$\thanks{Corresponding author}
\\[0.5em]
$^{1}$Audio, Speech and Language Processing Group (ASLP@NPU), School of Computer Science, Northwestern Polytechnical University\\
$^{2}$Yutu Zhineng\\
$^{3}$Shanghai Lingguang Zhaxian Technology\\
$^{4}$WeNet Open Source Community 
}

\newcommand{\tabincell}[2]{\begin{tabular}{@{}#1@{}}#2\end{tabular}}
\newcommand{\fix}{\marginpar{FIX}}
\newcommand{\new}{\marginpar{NEW}}
\newcommand{\method}{VoiceSculptor\xspace}
\newcommand{\jin}[1]{{\color{blue}#1}}

\maketitle
% \footnotetext{$^\dagger$ Corresponding author}

\begin{abstract}
% Despite significant advances in text-to-speech (TTS) technology, flexible and fine-grained control over speech attributes remains limited. Existing systems primarily rely on text and reference audio to generate speech, enabling speaker timbre imitation but offering little direct control over attributes such as pitch, speaking rate, age, emotional expression, speaking style, and others. We present VoiceSculptor, a unified system that integrates voice design and voice cloning. The voice design module generates speaker timbre from natural language descriptions, supports iterative instruction refinement via Retrieval-Augmented Generation (RAG), and enables fine-grained control over multiple voice attributes. The synthesized audio is then used as a prompt waveform for a voice-cloning model, enabling accurate speaker timbre cloning and supporting downstream speech synthesis tasks. To promote reproducibility and community research, we fully open-source our codebase and pretrained models.
Despite rapid progress in text-to-speech (TTS), open-source systems still lack truly instruction-following, fine-grained control over core speech attributes (e.g., pitch, speaking rate, age, emotion, style, and others). We present VoiceSculptor, an open-source unified system that bridges this gap by integrating instruction-based voice design and high-fidelity voice cloning in a single framework. It generates controllable speaker timbre directly from natural-language descriptions, supports iterative refinement via Retrieval-Augmented Generation (RAG), and provides attribute-level edits across multiple dimensions. The designed voice is then rendered into a prompt waveform and fed into a cloning model to enable high-fidelity timbre transfer for downstream speech synthesis. VoiceSculptor achieves open-source state-of-the-art (SOTA) on InstructTTSEval-Zh, and is fully open-sourced, including code and pretrained models, to advance reproducible instruction-controlled TTS research.
\end{abstract}

\begin{figure}[tbh]
    \centering
    \includegraphics[width=0.95\textwidth]{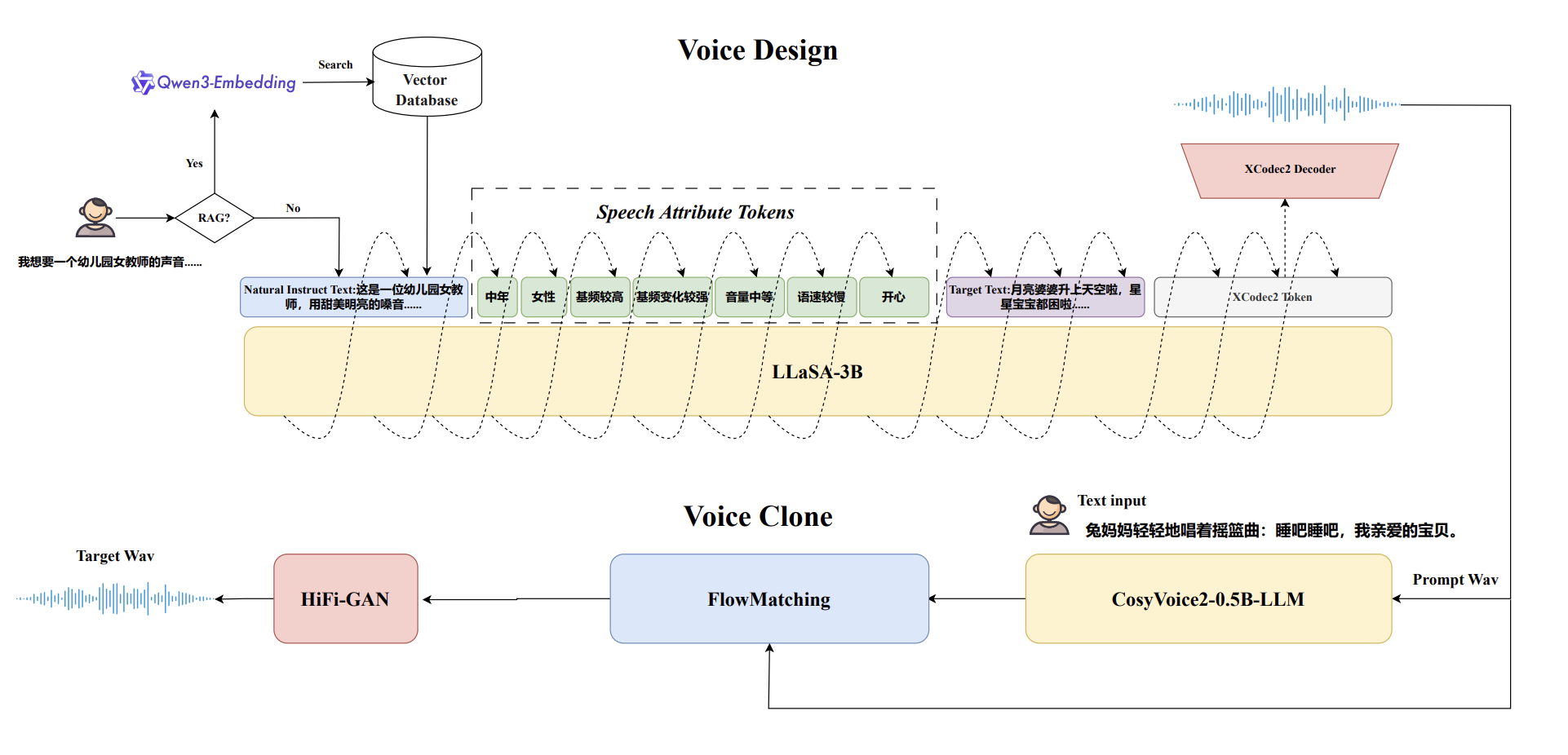}
    \caption{The overview of \method, which is composed of two core components: voice design and voice clone.}
    \label{fig: intro}
\end{figure}

\section{Introduction}
\label{sec:intro}
% 生成任务的大背景 和 生成式语音的背景
% In recent years, the rapid evolution of large-scale multimodal foundation models has fundamentally reshaped the paradigm of generative artificial intelligence (AI), enabling unified generation across text, speech, images, and videos. Commercial systems such as Gemini 2.5 Pro \& Flash, and GPT-4o mini have demonstrated strong instruction-following and multimodal reasoning capabilities. At the same time, recent text-to-video and audio–visual generation models, including Veo 3, Wan 2.6, Seedance 1.5 pro, and Kling 2.6, have shown impressive progress in jointly synthesizing coherent visual content and synchronized audio. These advances highlight a growing trend toward holistic, natural-language-driven content creation, in which users expect to express complex intentions once and receive rich, accurate multimodal outputs.

In recent years, the rapid evolution of large-scale multimodal foundation models has fundamentally reshaped the paradigm of generative artificial intelligence (AI), enabling unified generation across multiple modalities, including text, speech, images, and videos. Commercial systems such as Gemini 2.5 Pro, Gemini 2.5 Flash ~\citep{comanici2025gemini} and GPT-4o mini \footnote{\url{https://platform.openai.com/docs/models/gpt-4o-mini}} have demonstrated strong instruction-following and multimodal reasoning capabilities. In contrast, recent text-to-video and audio–visual generation models, including Veo 3 \footnote{\url{https://storage.googleapis.com/deepmind-media/veo/Veo-3-Tech-Report.pdf}}, Wan 2.6 \footnote{\url{https://www.wan-ai.co/wan-2-6}}, Seedance 1.5 Pro ~\citep{seedance2025seedance15pronative}, and Kling 2.6 \footnote{\url{https://app.klingai.com/global/}}, have shown impressive progress in jointly synthesizing coherent visual content with synchronized audio. Text-driven generation has become the dominant interface across modalities, with text-to-image, text-to-video, and video–audio joint generation systems all supporting natural-language descriptions as the primary control mechanism. 

At the same time, prior speech generation research has primarily focused on zero-shot voice synthesis, while paying comparatively less attention to fine-grained, instruction-driven, controllable speech generation from natural language. Modern neural text-to-speech (TTS) systems, including CosyVoice2 ~\citep{du2024cosyvoice}, LLaSA ~\citep{ye2025llasa}, F5-TTS ~\citep{chen2025f5}, SparkTTS ~\citep{wang2025spark}, and Index-TTS2 ~\citep{zhou2025indextts2}, can now generate highly natural speech and effectively mimic speaker timbre when reference audio is available. However, despite these advances, controllability over speech attributes remains limited, especially when compared to the flexibility observed in recent multimodal generation systems.

% instruct的技术发展
Beyond zero-shot capability, recent studies are increasingly exploring fine-grained, editable, and instructi-\\on-conditioned speech generation, aiming to provide users with more precise and expressive control over generated content across speech domains. To better understand this emerging line of research, it is essential to review how controllability has traditionally been achieved in speech synthesis. 

% traditional
% instruct "clip" based
Earlier controllable TTS methods typically relied on explicit prompt engineering or learned latent representations, rather than directly modeling instruction semantics from natural language, such as PromptStyle~\citep{liu2023promptstyle} and PromptSpeaker~\citep{zhang2023promptspeaker}. While these approaches laid the groundwork for controllable speech generation, their dependence on fixed templates and limited attribute spaces restricts scalability toward more flexible, instruction-driven voice design. Recent studies increasingly frame controllable speech generation as an alignment problem across cross-modal representations. UniStyle ~\citep{10.1145/3664647.3681465} unifies textual and acoustic style cues by mapping them—via a UniConnector—into a shared latent space, which is then conditioned by a VALL-E-style ~\citep{wang2023neuralcodeclanguagemodels} autoregressive decoder. This design allows either a short text prompt or a 3-second reference audio to guide prosody without additional adapters. FleSpeech ~\citep{li2025flespeech} extends this idea further, explicitly aligning text, audio, and visual captions within a unified latent space. Similarly, HiStyle ~\citep{zhang2025histyle} adopts a hierarchical design, stacking two lightweight Transformers that sequentially convert a text prompt into coarse speaker and fine-grained style embeddings. Despite their architectural differences, these approaches all compress rich voice attributes into a single continuous vector, leaving the downstream generator with only an implicit, entangled control signal. While effective, this design imposes an inherent limitation: the control input is low-bandwidth and lacks explicit factorization of acoustic attributes, leaving fine-grained, compositional control coarse and difficult to interpret. Recent work has begun to explore direct instruction understanding via large language models to achieve more interpretable, fine-grained, and controllable speech generation. For example, VoxInstruct ~\citep{zhou2024voxinstruct} unifies content and style into a single natural-language instruction and employs an MT5 encoder ~\citep{xue2021mt5massivelymultilingualpretrained} with a LLaMA-style decoder ~\citep{touvron2023llamaopenefficientfoundation} to generate discrete speech semantic tokens. Despite supporting free-form instructions, VoxInstruct compresses content and style into a single token sequence, limiting precise control over fine-grained prosodic attributes.

% 引出现存问题
Nevertheless, most existing systems still generate speech primarily conditioned on text and reference audio, offering limited direct control over fine-grained acoustic attributes such as pitch, speaking
rate, age, emotional expression, speaking style, and others. This gap reveals a fundamental bottleneck in current generative systems: although natural language has become the dominant interface for controlling complex multimodal generation, speech synthesis still lacks a principled, flexible mechanism for translating high-level linguistic intent into fine-grained acoustic realization. In contrast to visual and video generation, voice generation remains heavily constrained by reference-based conditioning or rigid control tokens.

% 基座理解模型
Recent advances in audio-centric foundation models, such as MiMo-Audio ~\citep{zhang2025mimo} and Step-Audio2 ~\citep{wu2025step}, demonstrate that scaling both model capacity and training data can substantially enhance representation quality and instruction-following capability in speech models. Their strong performance shows that large-scale models can better capture the semantics of textual instructions and generalize across speakers and styles, laying the groundwork for unified text-and-audio modeling. These developments suggest that LLM-like modeling with sufficient data and parameters is a promising approach to achieving fine-grained, natural-language-driven voice control, motivating the design of our method.

% 引出我们的工作
To address these limitations and build on insights from recent audio-centric foundation models, we propose \method. This unified and highly flexible speech synthesis framework bridges natural language intent and fine-grained voice generation. Unlike conventional TTS systems that rely solely on fixed control tokens or reference audio, VoiceSculptor enables users to design speaker timbre and manipulate multiple voice attributes directly via free-form natural language instructions. By providing both open-source accessibility and instruction-driven, fine-grained controllability, our method closes a key gap between existing open-source frameworks, which often lack expressiveness, and commercial closed-source systems, which offer limited transparency and reproducibility. Our method closes a key gap between existing open-source frameworks and commercial closed-source systems.

At the core of this capability, the voice design module introduces a chain-of-thought (CoT)-based~\citep{wei2022chain}, fine-grained attribute modeling mechanism that explicitly decomposes high-level natural language instructions into structured intermediate reasoning steps across multiple acoustic and stylistic attributes. By modeling this reasoning process as auxiliary attribute tokens, the model is guided to interpret abstract linguistic descriptions step by step and map them to concrete acoustic realizations, enabling precise, disentangled control over prosody, style, and speaker-related characteristics.

To further enhance instruction understanding and robustness, the voice design module incorporates Retrieval-Augmented Generation (RAG)~\citep{lewis2021retrievalaugmentedgenerationknowledgeintensivenlp}, which retrieves semantically relevant instruction examples and attribute knowledge to support iterative instruction refinement and generalization to out-of-domain descriptions. The framework integrates a voice design module with a voice cloning module, enabling synthesized audio from descriptive instructions to serve as a prompt waveform for downstream speech synthesis. By jointly leveraging CoT-based fine-grained attribute reasoning and RAG-based instruction grounding, VoiceSculptor establishes a more expressive, intuitive, and scalable paradigm for personalized, highly controllable TTS, aligning speech generation with the broader trajectory of multimodal generative systems.

\section{Architecture}
\begin{figure}[tbh]
    \centering
    \includegraphics[width=0.9\textwidth]{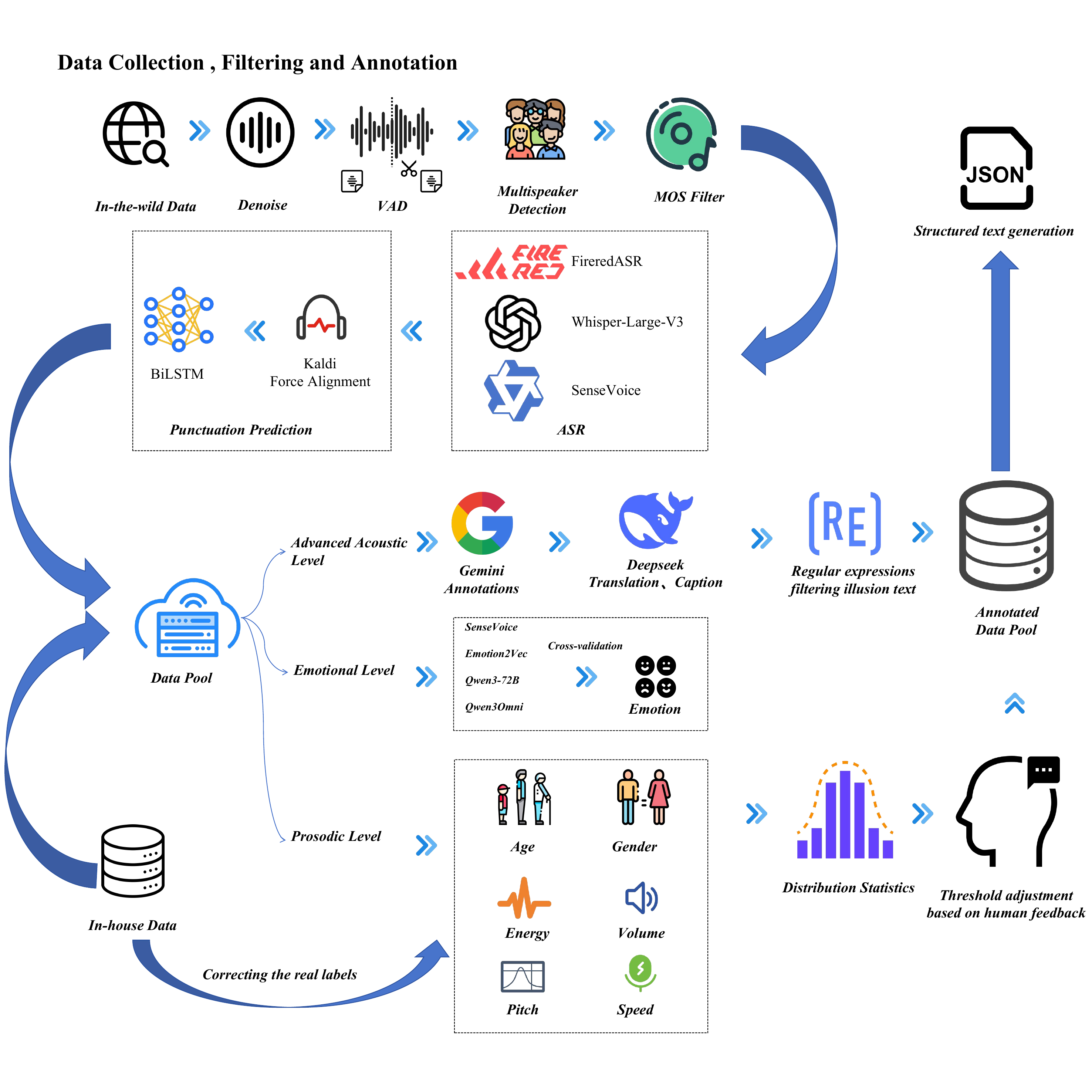}
    \caption{The data pipeline of building \method.}
    \label{fig:data_pipe}
\end{figure}

\subsection{Overview}

As shown in Figure~\ref{fig: intro}, \method adopts LLaSA-3B ~\citep{ye2025llasa} as the voice design model and CosyVoice2 ~\citep{du2024cosyvoice} as the voice cloning model. LLaSA is built upon the open-source LLaMA ~\citep{touvron2023llamaopenefficientfoundation} family released by Meta and is fine-tuned to leverage the strong text understanding and sequence modeling capabilities of large language models. To enable speech generation within an LLM framework, LLaSA incorporates an advanced neural audio codec, XCodec2, that converts continuous speech waveforms into discrete audio tokens resembling text tokens.

With the introduction of XCodec2, speech synthesis is reformulated as a sequence-to-sequence generation problem. Given a natural language instruction, the LLM is responsible for interpreting the semantic and stylistic intent of the text and predicting a corresponding sequence of discrete audio tokens. These predict the audio codec, then decode tokens to reconstruct high-quality speech waveforms. This design allows the model to jointly reason over text and speech tokens using a unified autoregressive generation paradigm.

Built on this foundation, \method comprises three key contributions. First, we construct a comprehensive data processing pipeline that supports large-scale data collection, filtering, multi-dimensional annotation, and human verification, providing high-quality supervision for natural-language-driven voice design. Second, we introduce CoT–based fine-grained attribute modeling, which enables precise, interpretable control over prosodic and stylistic attributes by explicitly guiding the LLM to reason over attribute-related semantics during audio token generation. Third, to improve robustness and generalization to diverse, out-of-domain natural language instructions, we incorporate a RAG mechanism that provides semantically aligned in-domain instruction examples at inference time, effectively grounding the instruction interpretation process.

Together, these components enable \method to translate natural language instructions into controllable, reusable voice representations that can be seamlessly consumed by downstream speech synthesis models such as CosyVoice2.

% Our main contributions are summarized as follows:
% \begin{itemize}
%     \item A comprehensive data processing pipeline. We design a complete data pipeline encompassing large-scale data processing, filtering, multi-dimensional annotation, and human verification, providing high-quality supervision for controllable speech synthesis.
%     \item Natural-language-driven voice control. We enable direct, flexible voice design through natural-language instructions by jointly modeling instruction text, structured attribute representations, and speech tokens, thereby significantly improving instruction understanding and controllability.
%     \item Fine-grained attribute controllability. We achieve precise control over multiple prosodic and stylistic attributes by introducing fine-grained attribute supervision and training strategies that encourage the model to rely on natural language semantics rather than explicit attribute tokens.
%     \item Retrieval-augmented instruction generalization. We incorporate retrieval-augmented generation to enhance robustness and generalization to out-of-domain natural language instructions by leveraging semantically similar instruction examples during inference.
% \end{itemize}

\subsection{Data Processing Pipeline}

The Figure~\ref{fig:data_pipe} illustrates the end-to-end data collection, filtering, annotation, and validation pipeline designed for VoiceSculptor. The pipeline starts from large-scale in-the-wild data and in-house data, which are first subjected to a series of automatic preprocessing steps, including denoising~\footnote{\url{https://github.com/Audio-WestlakeU/CleanMel}}, voice activity detection (VAD)~\footnote{\url{https://github.com/wiseman/py-webrtcvad}}, multi-speaker detection~\footnote{\url{https://github.com/pyannote/pyannote-audio}}, and perceptual quality filtering~\footnote{\url{https://github.com/AndreevP/wvmos}}, to ensure basic acoustic cleanliness and speaker consistency.

For linguistic alignment, automatic speech recognition (ASR) and forced alignment are applied to obtain accurate transcriptions and punctuation information. Specifically, FireRedASR ~\citep{xu2025fireredasropensourceindustrialgrademandarin} is used to transcribe Chinese audio, Whisper ~\citep{radford2022robustspeechrecognitionlargescale} is employed for English audio, and SenseVoice ~\citep{an2024funaudiollmvoiceunderstandinggeneration} is used for ASR cross-validation, as well as language and emotion recognition. The resulting transcripts are further processed using the Kaidi~\footnote{\url{https://github.com/sunsetsonwheels/kaidi}} alignment tool to perform character-level (for Chinese) or word-level forced alignment, producing precise timestamps and pause durations between adjacent tokens. Based on these fine-grained temporal cues, a trained punctuation-prediction model is applied to restore punctuation, yielding linguistically coherent and temporally aligned text–audio pairs. 

Based on the aligned text–audio representations, a structured annotation process is subsequently conducted across multiple levels. At the advanced acoustic level, each audio sample is first analyzed using Gemini 2.5 Pro ~\citep{comanici2025gemini} to obtain multi-dimensional annotations, including pitch, speaking rate, loudness, speaker gender and age, emotional state, paralinguistic characteristics, and contextual attributes. Based on these structured annotations, DeepSeek ~\citep{liu2025deepseek} is subsequently employed to perform translation and caption generation, producing natural-language descriptions of the audio style and vocal characteristics. To mitigate hallucinations introduced by large language models, we apply a set of rule-based regular expression filters to remove inconsistent or unsupported content. This process yields relatively high-quality, semantically grounded annotations and captions that serve as reliable supervision for natural-language-driven voice design. At the emotional level, speech is annotated through a cross-validated emotion labeling process. Specifically, we employ multiple complementary models, including Emo2Vec ~\citep{ma2024emotion2vec} , Qwen3-72B ~\citep{yang2025qwen3} , SenseVoice, and Qwen3-Omni ~\citep{xu2025qwen3omnitechnicalreport}, to independently predict emotion-related attributes from the audio. The outputs from these models are then cross-validated to resolve inconsistencies and improve label reliability, yielding a final set of emotion annotations with greater robustness and accuracy. At the prosodic level, fine-grained acoustic attributes are annotated through a combination of automatic estimation, statistical analysis, and human verification. Specifically, we employ the DataSpeech ~\citep{lyth2024naturallanguageguidancehighfidelity} model to estimate continuous prosodic features, including the mean and standard deviation of pitch, mean and standard deviation of energy, speaking rate, and loudness. Speaker gender and age are annotated using VoxProfile \citep{feng2025voxprofilespeechfoundationmodel}. Based on the extracted attributes, we perform distributional analysis across the dataset and conduct targeted human listening to calibrate attribute boundaries. The continuous prosodic features are discretized into 5 intervals per dimension, while age is categorized into 4 groups (child, youth, middle-aged, and elderly). Finally, for samples from in-house datasets, we leverage available ground-truth annotations to correct and refine the predicted age and gender labels, yielding a set of reliable, structured prosodic annotations.

All annotations are aggregated into a unified annotated data pool, where natural language descriptions are structured and regularized to form consistent instruction-like text representations. 

Through this comprehensive, iterative pipeline, VoiceSculptor constructs a high-quality, multidimensionally annotated dataset that supports natural-language-driven voice control, fine-grained attribute manipulation, and robust instruction understanding.

\subsection{CoT–based Fine-grained Attribute Modeling}

Various fine-grained acoustic attributes in speech signals—such as pitch, loudness, speaking rate, and temporal dynamics—play a critical role in shaping prosodic structure, expressive rhythm, and overall vocal style. However, directly conditioning speech synthesis models on explicit attribute tokens often leads to brittle control and over-reliance on structured inputs, limiting the models' ability to generalize to diverse natural-language instructions.

To address this challenge, we introduce a CoT–based fine-grained attribute modeling strategy that explicitly guides the model to reason over acoustic attributes through intermediate semantic representations. Instead of treating fine-grained attributes as independent control signals, CoT organizes attribute information into structured reasoning steps that bridge natural language instructions and acoustic realizations. This design enables the model to interpret high-level textual descriptions, decompose them into attribute-related semantics, and subsequently generate speech tokens that reflect the desired prosodic and stylistic characteristics.

Based on this formulation, VoiceSculptor jointly models instruction text, CoT-based fine-grained attribute tokens, and discrete speech tokens within a unified autoregressive framework. This allows the model to explicitly control multiple acoustic dimensions during synthesis, supporting precise prosody manipulation and flexible rendering of diverse vocal styles.

Moreover, to strengthen alignment between linguistic instruction and the generated speech representation, we adopt a joint cross-entropy (CE) training objective over both text and speech tokens. Specifically, the model is trained to predict not only the acoustic attribute tokens and intermediate linguistic tokens, but also the corresponding discrete speech tokens produced by the neural codec. By simultaneously optimizing CE loss across both modalities, the model learns a unified latent space that tightly couples natural-language semantics with low-level acoustic realization.

This multi-level supervision provides richer learning signals than training on speech tokens alone. In particular, the text-token CE loss encourages the model to capture semantic intent and attribute relationships better, while the speech-token CE loss guides accurate acoustic generation. Empirically, we observe that this joint optimization yields consistent performance improvements in instruction-following controllability, leading to more stable and precise attribute rendering in the synthesized speech.

% TODO:介绍drop的比例
Furthermore, to prevent over-reliance on explicit attribute tokens and to encourage a more profound understanding of instructions, we introduce a stochastic attribute token dropout strategy during training. Fine-grained attribute tokens are randomly removed from the input with a predefined probability, such as 0.2, forcing the model to infer the intended acoustic attributes solely from natural language instructions and contextual cues. This training strategy serves as an effective regularizer, improving the model's robustness and generalization in natural-language-driven voice control.

\subsection{Retrieval-augmented Instruction Generalization}
To improve the model's generalization ability and robustness when handling out-of-domain natural language instructions, we incorporate a RAG mechanism into the inference pipeline. This design enables the model to leverage prior knowledge encoded in semantically related in-domain instructions, thereby reducing sensitivity to distributional shifts in input prompts.

Specifically, we construct a vector-based instruction repository by embedding a large-scale collection of \textbf{500K in-domain natural language instructions}, which follow organizational and syntactic patterns similar to those observed in the training data, using the Qwen3-Embedding-0.6B model ~\citep{zhang2025qwen3embeddingadvancingtext}. The resulting high-dimensional semantic representations are stored in a Milvus~\footnote{\url{https://github.com/milvus-io/milvus}} vector database, enabling efficient large-scale similarity search during retrieval.

During inference, when retrieval augmentation is enabled, the incoming natural language instruction is first converted into a dense vector representation using the same embedding model. A cosine similarity–based semantic search is then performed against the vector database to identify the most relevant in-domain instructions. The retrieved instructions, semantically aligned with the input query, are subsequently injected into the model's input, guiding it toward a more stable and accurate interpretation of the user's intent.

By grounding the generation process in semantically similar in-domain examples, the proposed retrieval-augmented framework effectively mitigates the impact of unseen or structurally diverse instructions. This approach enhances both the robustness and controllability of the model under open-ended instruction scenarios, leading to more consistent and reliable generation performance across a wide range of natural language inputs.
\section{Experiments}
% 1. 探索了scaling对模型instruct能力的效果（对比llasa 1b、3b，对比1000h和3700h数据）
% 2. 文本端ce loss对最后指令跟随效果的影响（对比llasa 1b文本端是否加ce loss）
% 3. 探索了CoT对instruct跟随能力的影响（对比加tag和不加tag模型在效果上的差异）
% 4. rag+cosyvoice实现更好的指令跟随能力（对比rag的作用）

% 1. 对比实验
%   1. Minimax
%   2. 11labs
%   3. Gemini
%   4. gpt
%   5. cosyvoice2
%   6. llasa-instruct
%   7. cosyvoice2-instruct
% 2. scaling实验
%   1. 1000h llasa 1b
%   2. 1000h llasa 3b
%   3. 3700h llasa 1b
%   4. 3700h llasa 3b
% 3. CoT消融实验
%   1. 加tag做instruct（llasa 1b. 3700h）
%   2. 去掉tag做instruct（llasa 1b 3700h）
% 4. Ce loss消融实验
%   1. 加文本端ce loss（llasa 1b 1000h）
%   2. 去掉文本端ce loos（llasa 1b 1000h）
% 5. rag消融实验
%   1. Llasa
%   2. llasa+rag——》rag提供instruct文本（充当文本前端的角色）

% TODO:需要说明，rag库是通过llm（deepseek）构建了类似模板的文本数据库，然后通过qwen3emb模型进行语义检索，然后得到的文本指令

We conduct extensive experiments to comprehensively evaluate the effectiveness, controllability, and scalability of the proposed VoiceSculptor framework, with a particular focus on its voice design (VD) module. Given that our training data primarily consists of Chinese instruction–speech pairs, we adopt the InstructTTSEval-Zh benchmark ~\citep{huang2025instructttsevalbenchmarkingcomplexnaturallanguage} as the primary evaluation protocol to assess instruction-following performance in a controlled and fair setting. We compare VoiceSculptor against both strong open-source baselines and representative commercial systems, and further analyze its performance under different model sizes, data scales, and training strategies. In addition, we perform a series of ablation studies to isolate the contributions of key components, including CoT-based fine-grained attribute tokens, text-side cross-entropy supervision, and RAG. Human subjective evaluations are also conducted to complement automated metrics and to validate adherence to perceptual instructions. Overall, the experimental results consistently demonstrate that VoiceSculptor achieves state-of-the-art performance among open instruction-following TTS systems, while exhibiting robust scalability and strong controllability across a wide range of settings.
% TODO:加引用
\subsection{Evaluation on InstructTTSEval-Zh Benchmark}
Our training data predominantly consists of Chinese instruction–speech pairs. Therefore, we focus our evaluation on the model’s instruction-following and controllability performance in Chinese, enabling a fair and consistent assessment aligned with the training distribution. We emphasize that this choice does not imply any inherent limitation of the proposed method to Chinese. Instead, Chinese is adopted as a representative language to demonstrate the feasibility and effectiveness of our approach, which can be naturally extended to other languages given appropriate instruction data.

To this end, we employ InstructTTSEval-Zh, a Chinese instruction-based TTS evaluation benchmark designed to measure how well a model follows natural-language instructions in speech synthesis. The benchmark evaluates multiple aspects of instruction controllability, including Acoustic-Parameter Specification (APS), Descriptive-Style Directive (DSD), and Role-Play (RP). These metrics are computed by synthesizing speech from instruction prompts and assessing the generated audio using a unified evaluation protocol with a large language model as the evaluator.

\begin{table}[htbp]
\caption{Performance Comparison Across Different Models on InstructTTSEval-Zh Benchmark}
\centering
\begin{tabular}{lcccc}
\toprule
Model & APS$\uparrow$ (\%) & DSD$\uparrow$ (\%) & RP$\uparrow$ (\%) & AVG$\uparrow$ (\%) \\
\midrule
Gemini 2.5-Flash* ~\citep{comanici2025gemini} & 88.2 & 90.9 & 77.3 & 85.4 \\
Gemini 2.5-Pro* ~\citep{comanici2025gemini}  & 89.0 & 90.1 & 75.5 & 84.8 \\
GPT-4o-Mini-TTS*  & 54.9 & 52.3 & 46.0 & 51.1 \\
ElevenLabs*     & 42.8 & 50.9 & 59.1 & 50.9 \\
VoxInstruct ~\citep{zhou2024voxinstruct}       & 47.5 & 52.3 & 42.6 & 47.5 \\
MiMo-Audio-7B-Instruct ~\citep{zhang2025mimo} & 70.1 & \textbf{66.1} & 57.1 & 64.5 \\
\midrule
VoiceSculptor-VD     & 75.7 & 64.7 & \textbf{61.5} & \textbf{67.6} \\
VoiceSculptor-VD \& VC & \textbf{77.2} & 65.1 & 59.6 & 67.3 \\ 
\bottomrule
\end{tabular}
\label{tab:model_comp}
\\[5pt]
\small * indicates commercial models, while the others are open-source. All metrics follow InstructTTSEval. For ElevenLabs and MiMo-Audio-7B-Instruct, speech samples are generated using either the official APIs or the released open-source models, and evaluated with Gemini 2.5 Pro, from which the reported results of the compared models are obtained.
\end{table}

As shown in the table ~\ref{tab:model_comp}, all results of VoiceSculptor are reported with RAG enabled, where an external instruction text repository provides semantically aligned in-domain guidance during inference, ensuring stable and robust instruction grounding for the voice design module.

In this evaluation setting, VoiceSculptor-VD consistently surpasses all open-source and instruction-tuned baselines across most evaluation metrics. Notably, it attains the best performance on APS and RP, reflecting its superior proficiency in both attribute perception and role-play within the voice design module. These findings indicate that, when integrated with retrieval-based instruction grounding, the VD module of VoiceSculptor is highly effective at mapping abstract natural language descriptions to precise and controllable vocal characteristics, which aligns with the fundamental goal of voice design.

Although Gemini 2.5-Flash and Gemini 2.5-Pro achieve strong overall scores, they are commercial, proprietary models that rely on closed-source infrastructure. In contrast, VoiceSculptor achieves state-of-the-art (SOTA) performance among open-source instruction-following TTS models, while maintaining a unified, natural-language-driven voice design framework.

Compared with MiMo-Audio-7B-Instruct, which achieves slightly higher performance on DSD, VoiceScul-\\ptor-VD shows clear advantages in APS and RP, leading to the highest overall AVG score among open systems. This indicates that the voice design module of VoiceSculptor prioritizes precise and consistent attribute control over isolated stylistic similarity, resulting in more faithful and reliable instruction-following behavior at the voice design stage. The substantial performance margin over VoxInstruct and GPT-4o-Mini-TTS further highlights the effectiveness of the design choices specific to the VD module, including fine-grained attribute modeling, instruction-aware training, and robust scaling strategies.

To further account for the requirements of downstream speech synthesis tasks, particularly the preservation of stylistic characteristics conveyed by prompt waveforms, we additionally report results for VoiceSculptor-VD and VC. In this setting, the prompt waveform generated by the voice design module, along with the corresponding test text, is fed into the CosyVoice2 model for downstream speech synthesis. This evaluation protocol allows us to directly assess whether the style specified and generated at the voice design stage can be faithfully retained when transferred to a subsequent synthesis model. Experimental results indicate that the stylistic attributes encoded in the prompt waveforms produced by VoiceSculptor-VD are preserved mainly in the downstream synthesized speech. Despite the introduction of an additional synthesis stage and a different model architecture, the generated speech maintains strong consistency with the intended characteristics, demonstrating effective style transfer and robustness of the designed prompt representations. These findings suggest that the voice design outputs of VoiceSculptor-VD are not only effective in isolation but also serve as reliable, reusable conditioning signals for downstream text-to-speech systems, thereby supporting practical deployment scenarios in which voice design and speech synthesis are decoupled.

Overall, these results demonstrate that the voice design module of VoiceSculptor achieves state-of-the-art performance among instruction-following TTS systems on the voice design benchmark. The consistent improvements across multiple complementary metrics confirm that the VD module can accurately interpret natural-language instructions and reliably generate the desired vocal attributes, establishing VoiceSculptor’s voice design component as a strong, practical solution for controllable voice design in natural-language-driven text-to-speech systems.

\subsection{Scaling Study on Model Size and Data Size}

Evaluating instruction-following performance using large-scale automated benchmarks requires repeated calls to proprietary models such as Gemini, which significantly increases evaluation cost and latency. To enable rapid, iterative validation of scaling trends, we construct an internal, lightweight subjective evaluation benchmark comprising 100 carefully curated test instructions and their corresponding synthesized speech outputs.

\begin{table}[htbp]
\centering
\caption{Overview of Training Data Composition}
\label{tab:data_description}
\begin{tabular}{lc p{8cm}}
\toprule
Dataset Size & Dataset Name & Data Sources \\
\midrule
1{,}000 h  & $SFT_{Data1}$ &
In-the-wild data. \\

3{,}700 h  & $SFT_{Data2}$ &
Combination of in-the-wild data and internal data. \\

4{,}000 h  & $SFT_{Data3}$ &
Combination of more in-the-wild data and internal data. \\

9{,}000 h  & $CPT_{Data4}$ &
Combination of more in-the-wild data , emotion-filtered samples from the open-source \textit{VoxBox} ~\citep{wang2025spark} dataset, and internal data. \\
\bottomrule
\end{tabular}
\end{table}

\begin{table}[htbp]
\centering
\caption{Scaling Study on Model Size and Data Size}
\label{tab:scaling_study}
\begin{tabularx}{\linewidth}{l c c *{5}{>{\centering\arraybackslash}X}}
\toprule
\makecell{Model\\Size} & \makecell{Training\\Data} & \makecell{Epoch} 
& IMOS$\uparrow$ & APS$\uparrow$(\%) & DSD$\uparrow$(\%) 
& RP$\uparrow$(\%) & AVG$\uparrow$(\%) \\
\midrule
1B & $SFT_{Data1}$ & 3 epoch for SFT & 3.09$\pm$0.21 & 51.3 & 48.2 & 35.7 & 45.1 \\
3B & $SFT_{Data1}$ & 3 epoch for SFT & 3.24$\pm$0.18 & 59.2 & 53.1 & 39.4 & 50.6 \\
1B & $SFT_{Data2}$ & 3 epoch for SFT & 3.35$\pm$0.11 & 61.5 & 55.9 & 45.9 & 54.4 \\
3B & $SFT_{Data2}$ & 3 epoch for SFT & 3.58$\pm$0.24 & 72.4 & 60.0 & 52.8 & 61.8 \\
3B & \makecell{$CPT_{Data4}$\\and $SFT_{Data3}$}
& \makecell{2 epoch for CPT\\and 3 epoch for SFT}
& \textbf{3.67$\pm$0.13} & \textbf{75.7} & \textbf{64.7} & \textbf{61.5} & \textbf{67.6} \\
\bottomrule
\end{tabularx}
\end{table}

We conduct a human listening study focusing on instruction-following capability, measured by Instructi-\\on-following Mean Opinion Score (IMOS). IMOS assesses explicitly how well synthesized speech adheres to the semantic and stylistic requirements expressed in the natural-language instruction, rather than overall audio quality alone.
A total of 33 human listeners participated in the evaluation. Each listener is randomly assigned 10 audio samples drawn from the 100-sample test set, ensuring that each test sample is evaluated by multiple listeners while avoiding listener fatigue. Listeners are asked to provide subjective scores based on how accurately the synthesized speech follows the given instruction, using a standardized MOS-style rating scale. The final IMOS score is computed by averaging all listener ratings across samples.

\begin{figure}[tbh]
    \centering
    \includegraphics[width=0.8\textwidth]{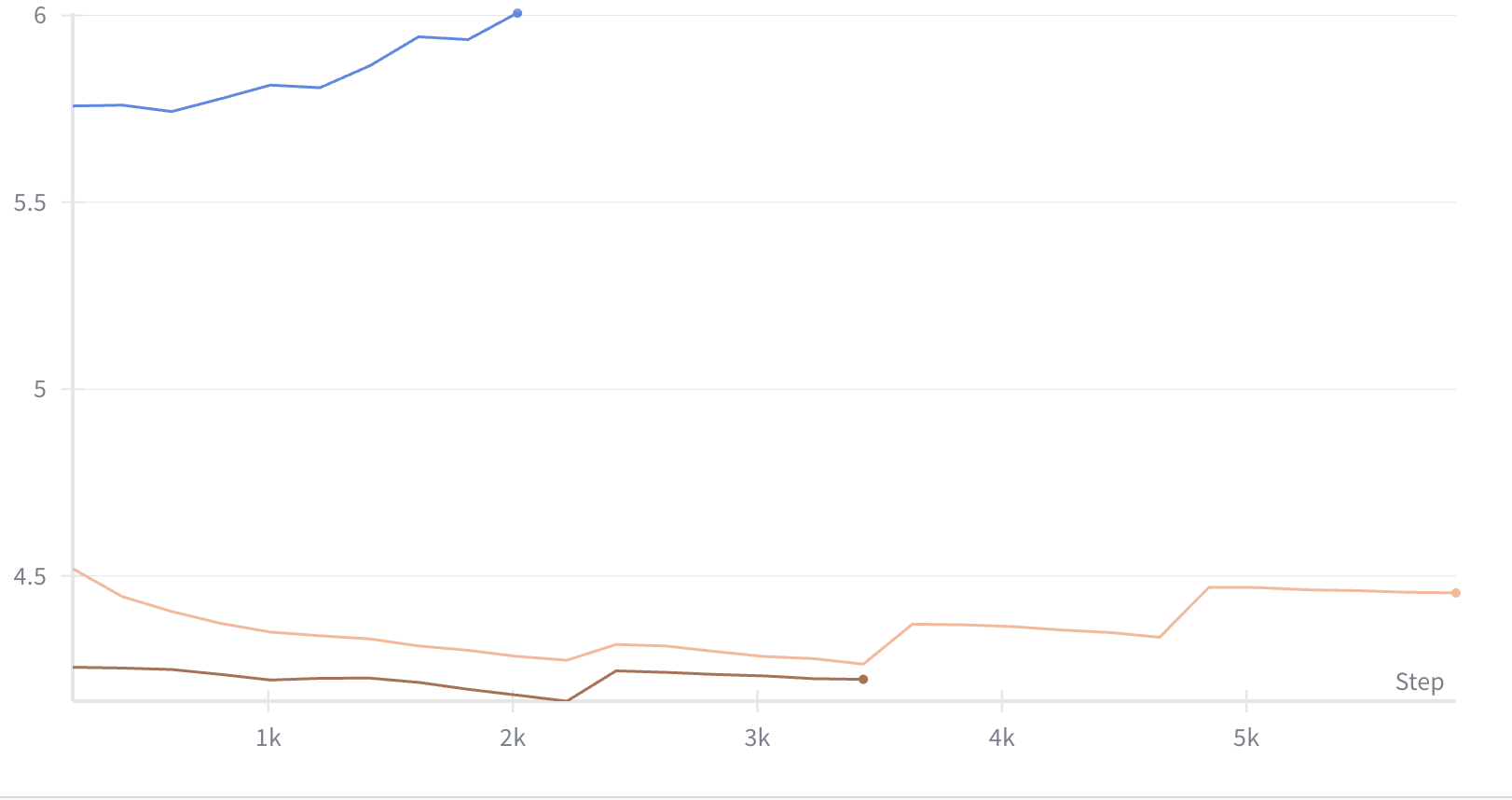}
    \caption{Validation loss curves of the SFT stage under different data configurations.}
    \label{fig:scaling-data}
\end{figure}

Table~\ref{tab:data_description} further details the duration and source composition of each training dataset, providing a clearer view of how data scale and diversity evolve across different training stages. The results in table ~\ref{tab:scaling_study} demonstrate that model performance consistently benefits from both increased model capacity and enlarged, more diverse training data. Under identical Supervised Fine-Tuning (SFT) settings, scaling the model from 1B to 3B parameters yields clear improvements across all metrics, indicating stronger representation and generalization capabilities. For a fixed model size, expanding SFT data from in-the-wild to larger mixtures with internal data yields substantial gains, highlighting the importance of data scale and diversity. Finally, incorporating large-scale emotion-aware continual pretraining (CPT) on $CPT_{Data4}$ before SFT achieves the best overall performance, suggesting that CPT provides a more favorable initialization and enables the model to better exploit downstream supervised data for the voice design task. Figure~\ref{fig:scaling-CoT} and Figure~\ref{fig:scaling-data} further corroborate these findings by showing consistently lower validation loss for larger models, and richer training data throughout training.

Increasing model capacity and enriching the quality and diversity of training data are fundamental to improving speech instruction-following. Building on this, the CPT stage offers a more robust initialization for the SFT stage, thereby significantly boosting overall performance on the Voice Design task.

\subsection{Ablation Study on CoT-based Fine-grained Attribute Tokens}

Table~\ref{tab: CoT_ablation} presents the ablation results of the proposed CoT-based fine-grained attribute tokens on the InstructTTSEval-Zh evaluation benchmark. Incorporating CoT leads to consistent and substantial improvements across all evaluation metrics, including IMOS, APS, DSD, RP, and the overall AVG score, demonstrating the effectiveness of explicit chain-of-thought modeling for voice design tasks. These gains are observed without altering the model architecture, indicating that CoT primarily enhances controllability and attribute understanding rather than relying on increased model complexity.

\begin{table}[htbp]
\centering
\caption{Ablation Study of CoT-based fine-grained attribute tokens}
\label{tab: CoT_ablation}
\begin{tabular}{lccccc}
\toprule
Setting & IMOS$\uparrow$ & APS$\uparrow$ (\%) & DSD$\uparrow$ (\%) & RP$\uparrow$ (\%) & AVG$\uparrow$ (\%) \\
\midrule
VoiceSculptor-VD & \textbf{3.67$\pm$0.17} & \textbf{75.7} & \textbf{64.7} & \textbf{61.5} & \textbf{67.6} \\
VoiceSculptor-VD w/o CoT & 3.59$\pm$0.14 & 71.6 & 61.9 & 58.9 & 63.5 \\

\bottomrule
\end{tabular}
\end{table}

\begin{figure}[tbh]
    \centering
    \includegraphics[width=0.8\textwidth]{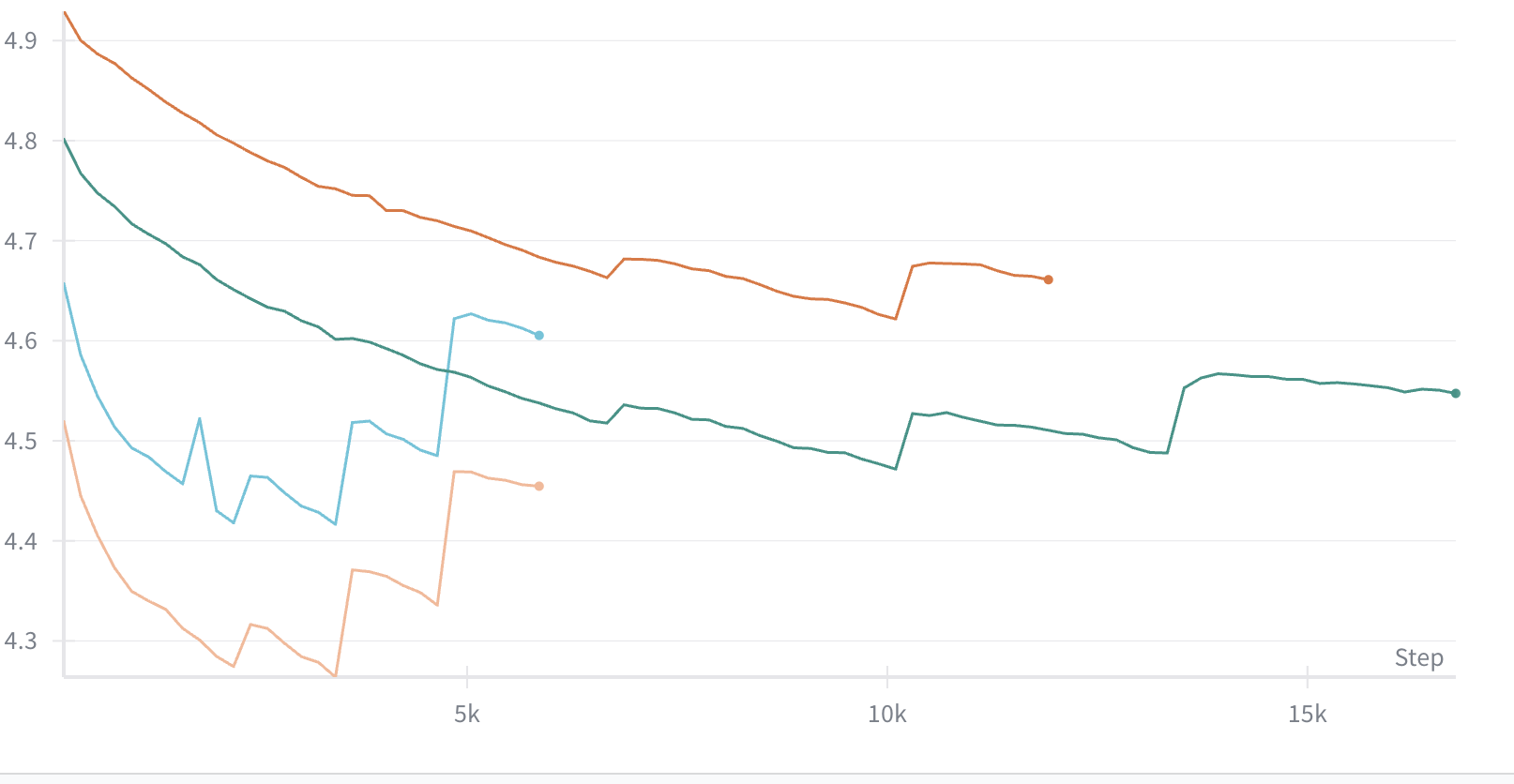}
    \caption{Validation loss curves for the ablation study on CoT and model scale.}
    \label{fig:scaling-CoT}
\end{figure}

Across both model scales, models equipped with CoT-based attribute tokens achieve better overall performance, suggesting that the proposed approach generalizes well to different model capacities. Figure~\ref{fig:scaling-CoT} provides supporting evidence from validation loss trends, showing that CoT-enhanced models exhibit more favorable optimization behavior during training.

Moreover, randomly dropping each attribute token with a probability of 0.2 during training does not degrade performance. On the contrary, the consistent improvements observed in Table~\ref{tab: CoT_ablation} indicate that this stochastic token dropout strategy serves as an effective regularization mechanism, encouraging the model to robustly integrate attribute information without overfitting to specific tokens.

The ablation study on CoT-based fine-grained attribute tokens validates the effectiveness of introducing auxiliary attribute tokens into the voice design framework. As shown by consistent improvements across all evaluation metrics, these tokens enable the model to capture better and utilize fine-grained attribute information, leading to more accurate and controllable speech generation. Beyond quantitative gains, the use of CoT-based attribute tokens facilitates finer-grained control over voice characteristics, allowing more precise manipulation of individual attributes. This enhanced controllability is further demonstrated through qualitative examples on our demo page, where the impact of fine-grained attribute conditioning is clearly evident.

\subsection{Ablation Study of Text Cross-Entropy Loss}

Based on the results in Table~\ref{tab:celoss_ablation}, introducing the text-side cross-entropy (CE) loss during training is highly effective. By jointly modeling the instruction text and audio tokens within a unified training objective, the model achieves consistent and significant improvements across all evaluation metrics. Compared to the setting without text CE loss, incorporating text CE loss yields notable gains in IMOS, APS, DSD, RP, and the overall AVG score, indicating greater alignment between textual instructions and generated speech.

\begin{table}[htbp]
\centering
\caption{Ablation Study Of Text Cross Entropy Loss}
\label{tab:celoss_ablation}
\begin{tabular}{lccccc}
\toprule
Setting & IMOS$\uparrow$ & APS$\uparrow$ (\%) & DSD$\uparrow$ (\%) & RP$\uparrow$ (\%) & AVG$\uparrow$ (\%) \\
\midrule
VoiceSculptor-VD & \textbf{3.67$\pm$0.15} & \textbf{75.7} & \textbf{64.7} & \textbf{61.5} & \textbf{67.6} \\
VoiceSculptor-VD w/o Text CE Loss & 3.42$\pm$0.23 & 67.9  & 59.4  &  58.2 & 61.8 \\
\bottomrule
\end{tabular}
\end{table}

These results suggest that explicitly supervising the text modality encourages the model to capture better long-range contextual dependencies and the semantic intent conveyed by instructions, rather than treating text merely as auxiliary conditioning. As a result, the model develops stronger context understanding and instruction-following capabilities, which directly translate into improved controllability and perceptual quality in the voice design task.

\subsection{Ablation Study of RAG}
%TODO 是否写到方法部分
% The retrieval memory used in our RAG framework is constructed from a large set of synthetic instruction texts. Specifically, we generate 500K instruction-style texts using DeepSeek, where the instructions are organized to closely match the format and distribution of the training instructions seen during supervised fine-tuning. To enable efficient retrieval, all instructions are encoded into dense text embeddings using Qwen3-0.6B. During inference, we retrieve the most relevant instructions based on cosine similarity between the query instruction and the embedding index, and then provide the retrieved texts to the model as auxiliary contextual information.

Table~\ref{tab:rag_ablation} demonstrates the necessity and effectiveness of incorporating RAG into the proposed VoiceSculpt-\\or-VD framework. Enabling RAG leads to substantial improvements across all evaluation metrics, with especially pronounced gains in APS (+7.1\%), RP (+13.0\%), and the overall AVG score (+8.2\%). These results indicate that external retrieval provides critical contextual and attribute-related information that the model alone may not reliably infer from the instruction text, thereby significantly enhancing controllability and instruction adherence in the voice design task.

\begin{table}[htbp]
\centering
\caption{Ablation Study of RAG}
\label{tab:rag_ablation}
\begin{tabular}{lccccc}
\toprule
Setting & IMOS$\uparrow$ & APS$\uparrow$ (\%) & DSD$\uparrow$ (\%) & RP$\uparrow$ (\%) & AVG$\uparrow$ (\%) \\
\midrule
VoiceSculptor-VD &\textbf{3.67$\pm$0.27} & \textbf{75.7} & \textbf{64.7} & \textbf{61.5} & \textbf{67.6} \\
VoiceSculptor-VD w/o RAG & 3.39$\pm$0.23 & 68.6 & 61.1 & 48.5 & 59.4 \\

% \midrule
% $\Delta$ (w/ RAG -- w/o RAG) & +0.28 & +7.1 & +3.6 & +13.0 & +8.2 \\
\bottomrule
\end{tabular}
\end{table}

At the same time, the significant performance gap between the RAG and non-RAG settings also reveals inherent limitations of the current model. In particular, the model exhibits relatively limited text understanding and generalization when relying solely on its internal representations, making it sensitive to instruction phrasing and less robust to unseen or complex textual descriptions. By supplementing the model with retrieved examples and structured attribute information, RAG effectively compensates for these shortcomings.

\section{Limitations and Future Work}
The current model still exhibits several limitations in practical use. First, the model may exhibit limited stability, where repeated synthesis under the same instruction occasionally fails to maintain precise control over the desired attributes. Second, during speech synthesis and interaction, the system occasionally suffers from long periods of silence or delayed responses, which negatively affect the user experience. In terms of training data coverage, the model does not yet perform well for elderly and child voices, and the naturalness and timbre consistency for these groups remain insufficient.

Moreover, our current evaluation of the model’s instruction-following ability is incomplete. Existing benchmarks focus primarily on Chinese tasks, and we have not yet conducted a comprehensive assessment of their English or multilingual capabilities.

In future work, we plan to address these limitations more fundamentally by strengthening the model’s text understanding capabilities. Specifically, we will explore incorporating large-scale text data during the pre-training stage to preserve better and enhance linguistic representations, performing instruction data augmentation to improve robustness and generalization, and adopting more semantically expressive audio representations by replacing the current xcodec2. In addition, we will introduce larger-scale and more diverse data to further improve the model’s adaptability and expressiveness across different voice characteristics. These improvements are expected to reduce the model’s reliance on external retrieval while maintaining strong instruction-following performance.

\section{Conclusion}
In this work, we present VoiceSculptor, a unified natural-language-driven voice design framework that enables fine-grained, controllable speech synthesis. Extensive evaluations on the InstructTTSEval-Zh benchmark demonstrate that VoiceSculptor-VD achieves SOTA performance among open-source instruction-following TTS systems, consistently outperforming strong baselines across multiple complementary metrics. Our scaling study confirms that instruction-following capability benefits predictably from increased model capacity, richer training data, and staged training strategies with large-scale continual pre-training. Through systematic ablation studies, we further validate the effectiveness of key design choices, including CoT-based fine-grained attribute tokens, text-side cross-entropy supervision, and retrieval-augmented generation, each of which contributes to improved instruction understanding, controllability, and robustness. In particular, CoT-based attribute modeling enables more precise and interpretable control over vocal characteristics. At the same time, RAG effectively compensates for the model’s limited text generalization by providing semantically aligned in-domain guidance. Finally, we demonstrate that the voice design outputs of VoiceSculptor can be reliably transferred to downstream speech synthesis models, supporting practical deployment scenarios where voice design and speech generation are decoupled. Together, these results establish VoiceSculptor as a scalable and effective solution for natural-language-driven voice design.

\section{Ethics Statement}
Do not use this model for unauthorized voice cloning, impersonation, fraud, scams, deepfakes, or any illegal or malicious activities.
Ensure compliance with local laws and regulations when using this model and uphold ethical standards.
The developers assume no liability for any misuse of this model.
Necessary clarification regarding generated voices:
As a generative model, the voices produced by this system are synthetic outputs inferred by the model, not recordings of authentic human voices.
The generated voice characteristics do not represent or reproduce any specific real individual, and are not derived from or intended to imitate identifiable persons.
We advocate for the responsible development and use of AI and encourage the community to uphold safety and ethical principles in AI research and applications.
\clearpage
% 下面的coding和多语言的细节，考虑放到附录。上面的总表其实也已经有所体现
\section{Appendix}
\label{sec:appendix}
For each model scale, the curves with lower validation loss correspond to models equipped with CoT-based fine-grained attribute tokens, while the curves with higher validation loss denote the baseline models without CoT attribute tokens, where each attribute token is randomly dropped with a probability of 0.2 during training.The upper orange/green curves represent the 1B model trained on 8×L40 GPUs, and the lower light-blue/light-orange curves represent the 3B model trained on 8×A100 GPUs.All curves report validation loss as a function of training steps.Our group's ablation experiments were validated using 3700 hours of data.We selected the epoch with the lowest validation set loss, i.e., epoch 3, as the final model for each of our experiments.

All models share the same 3B parameter architecture and are trained using identical optimization and training settings on 8×A100 GPUs. The only difference lies in the training data scale and pretraining strategy.From top to bottom, the curves correspond to 1000h SFT, 3700h SFT, and 9k h continued pretraining (CPT) followed by 3700h SFT, respectively.Consistent with previous observations, increasing the amount of supervised data reduces validation loss, while large-scale CPT further improves convergence and generalization, yielding the lowest loss throughout the SFT stage.

\clearpage
\bibliography{biblio}
\bibliographystyle{colm2024_conference}

\end{document}